\documentclass{aa}
\usepackage{txfonts}
\usepackage{graphicx}
\usepackage{subcaption}
\usepackage[colorlinks=true,allcolors=blue]{hyperref}

\usepackage{gensymb}
\begin{document} 

\authorrunning{Mountrichas et al.}
\titlerunning{AGN populations in the local universe}

%\title{AGN populations in the local universe: their position relative to the main-sequence, their stellar populations and accretion efficiency and the role of the AGN feedback}
\title{AGN populations in the local universe: their alignment with the main-sequence, characteristics of their stellar populations, accretion efficiency, and the impact of AGN feedback}

\author{G. Mountrichas\inst{1}, A. Ruiz\inst{2}, I. Georgantopoulos\inst{2}, E. Pouliasis\inst{2}, A. Akylas\inst{2} and E. Drigga\inst{2}}
          
     \institute {Instituto de Fisica de Cantabria (CSIC-Universidad de Cantabria), Avenida de los Castros, 39005 Santander, Spain
              \email{gmountrichas@gmail.com}
              \and
               National Observatory of Athens, Institute for Astronomy, Astrophysics, Space Applications and Remote Sensing, Ioannou Metaxa
and Vasileos Pavlou GR-15236, Athens, Greece}

\abstract{In this study, we utilize a sample of 338 galaxies within the redshift range of $\rm 0.02<z<0.1$, drawn from the Sloan Digital Sky Survey (SDSS), for which there are available classifications, based on their emission line ratios. We, further, identify and select Compton-thick (CT) AGN through the use of X-ray and infrared luminosities at $12\,\mu m$. We construct the spectral energy distributions (SEDs) for all sources and fit them using the CIGALE code to derive properties related to both the AGN and host galaxies. Employing stringent criteria to ensure the reliability of SED measurements, our final sample comprises 14 CT AGN, 118 Seyfert 2 (Sy2), 82 composite, and 124 LINER galaxies. Our analysis reveals that, irrespective of their classification, the majority of the sources lie below the star-forming main-sequence (MS). Additionally, a lower level of AGN activity is associated with a closer positioning to the MS. Utilizing the D$_n$4000 spectral index as a proxy for the age of stellar populations, we observe that LINERs exhibit the oldest stellar populations compared to other AGN classes. Conversely, CT sources are situated in galaxies with the youngest stellar populations. Furthermore, LINER and composite galaxies tend to show the lowest accretion efficiency, while CT AGN, on average, display the most efficient accretion among the four AGN populations. Our findings are consistent with a scenario in which the different AGN populations might not originate from the same AGN activity burst. Early triggers in gas rich environments can create high accretion rate SMBHs leading to a progression from CT to Sy2, while later triggers in gas poor stages result in low accretion rate SMBHs like those found in LINERs.}

   
\maketitle

\begin{figure*}
\centering
  \includegraphics[width=\textwidth]{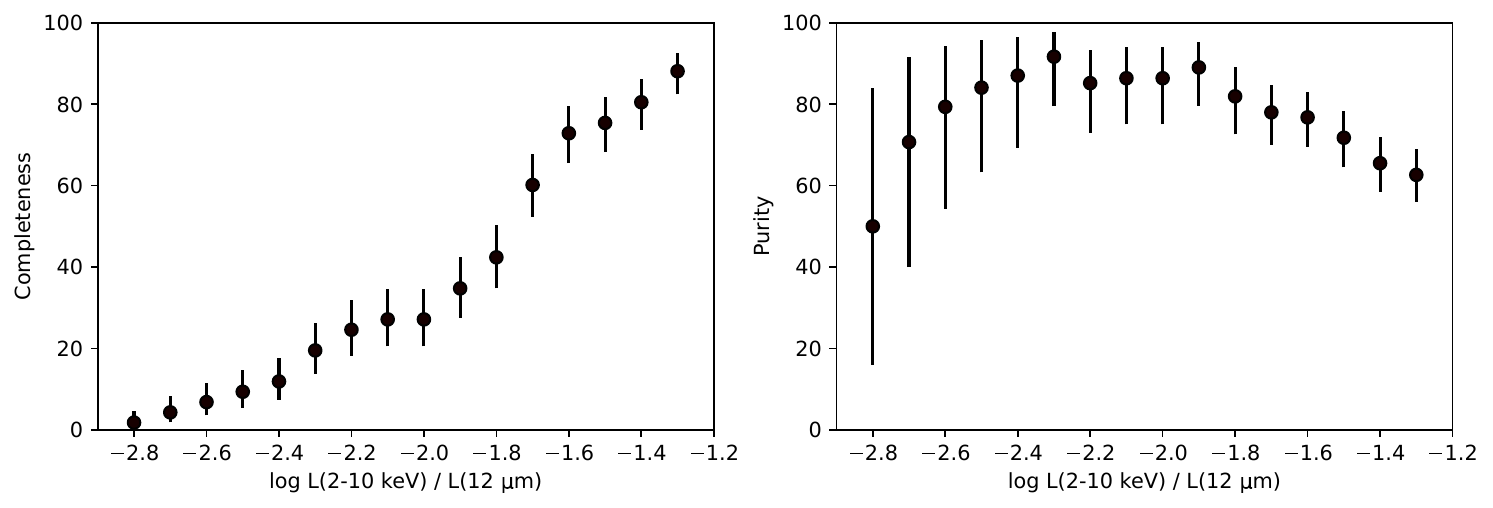} 
  \caption{Completeness (left) and purity (right) for CT AGN in the BASS sample depending on different selection criteria based on X-ray-to-MIR-luminosity ratio.}
  \label{fig_ct_diagnostic}
\end{figure*}  

\begin{figure}
\centering
 \includegraphics[width=\columnwidth]{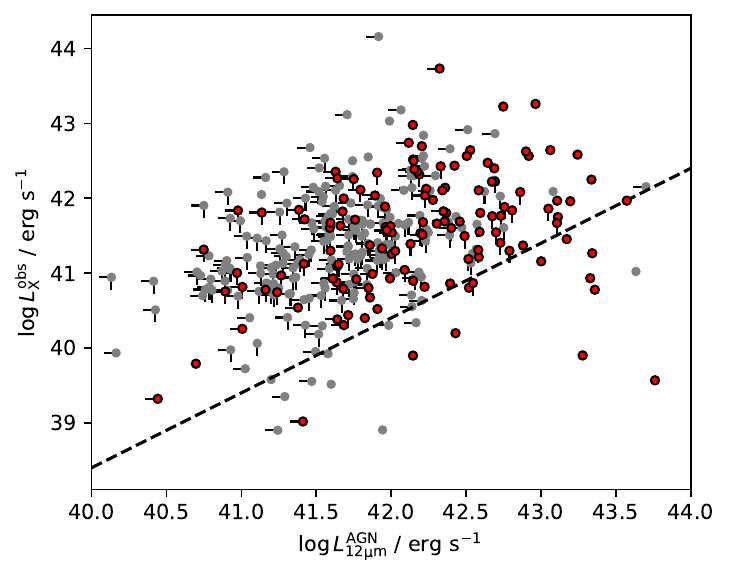} 
  \caption{AGN luminosity at $12~\mu m$, as estimated by CIGALE, versus observed X-ray luminosity in the 2-10~keV band for our sample of local SDSS AGN. Grey circles correspond to LINER/Composite objects, red circles are Seyfert 2 galaxies. Symbols marked with a horizontal and/or vertical bar show upper-limits in the $12~\mu m$ and/or X-ray luminosity, respectively. The black, dashed line shows $\log(L_\mathrm{X}^\mathrm{obs} / L_{12~\mu\mathrm{m}}^\mathrm{SGN}) = -1.6$.}
  \label{fig_ct_selection}
\end{figure}

\section{Introduction}

In the realm of active galactic nuclei (AGN), the local Universe presents a diverse population of galaxies hosting different AGN classes, each characterized by distinct observational features. Among these classes, there are LINERs (Low-Ionization Nuclear Emission-line Regions), Seyferts, and composite galaxies that consist of both AGN and star-forming systems, each offering valuable insights into the intricate interplay between supermassive black holes (SMBHs) and their host galaxies. Understanding the properties and behaviours of these AGN classes is crucial for unraveling the mechanisms that drive their activity and influence the evolution of their host systems.

Investigating the star formation rate (SFR) and stellar mass, M$_*$ of galaxies hosting AGN provides a crucial contextual framework for comprehending their evolutionary trajectories \citep[e.g.,][]{ROsario2012, Santini2012, Rosario2013, Mullaney2015, Masoura2018, Bernhard2019, Mountrichas2021b, Mountrichas2021c, Koutoulidis2022, Pouliasis2022, Mountrichas2023c, Mountrichas2023d, Mountrichas2024c}. These parameters offer essential clues about the ongoing astrophysical processes within these galaxies, shedding light on the co-evolution of AGN and their host galaxies. LINER, Seyferts and composite galaxies, serve as unique laboratories to explore the inter-dependencies between AGN activity, star formation, and the overall stellar content of their host systems.

The study of stellar populations across the different AGN populations offers a glimpse into the historical star formation activity within these systems \citep[e.g.,][]{Kauffmann2003, Kauffmann2003a, Kewley2006, Mountrichas2022c, Georgantopoulos2023}. Utilizing parameters like the D$_n$4000 spectral index, which serves as a proxy for the age of stellar populations, enables researchers to unravel the past evolutionary paths of these galaxies.

The Eddington ratio, denoted as n$_{Edd}$, emerges as a pivotal parameter in quantifying the accretion efficiency of AGN. Defined as the ratio of the bolometric luminosity to the Eddington luminosity (i.e., the maximum luminosity an AGN can emit; $\rm L_{Edd}=1.26\times 10^{38}\,M_{BH}/M_\odot\, erg\,s^{-1}$, where M$_{BH}$ is the mass of the SMBH), n$_{Edd}$ provides insights into the balance between radiation pressure and gravitational forces around the SMBH. Examining the n$_{Edd}$ values for different AGN populations allows for a comparative analysis of thThe evidence indicates that the different types of AGN we've examined may result from distinct phases of AGN activity. For example, if a supermassive black hole (SMBH) becomes active early in a galaxy's evolution, when there's plenty of gas, it might exhibit a high accretion rate, appearing as a Seyfert 2 (Sy2) and potentially transitioning from Compton-thick (CT) to Sy2. On the other hand, if the AGN activity begins later in the galaxy's timeline when gas is less abundant, the SMBH would likely have a lower accretion rate, leading to a Low-Ionization Nuclear Emission-line Region (LINER).eir accretion processes, offering a deeper understanding of the diverse ways in which AGN interact with their environments \citep[e.g.,][]{Kewley2006, Georgantopoulos2023, Mountrichas2024a, Mountrichas2024b}.

Previous studies have found that LINER galaxies tend to have higher M$_*$, redder optical colors and higher black hole masses compared to Seyferts \cite[e.g.][]{Smolcic2009}. LINERs also are more dusty and more concentrated than Seyferts, although, these differences could be, mainly, due to the different n$_{Edd}$ of the AGN populations, with LINERs to be dominant at low n$_{Edd}$ and Seyferts dominant at high n$_{Edd}$ \citep[e.g.,][]{Kewley2006}. Seyferts also appear to reside in dark matter holes with lower mass compared to LINERs \citep{Constantin2006}. For an in-depth overview of the various AGN populations, refer to \cite{Heckman2014}.

In this work, we use galaxies from the Sloan Digital Sky Survey \citep[SDSS;][]{Almeida2023} for which there are available classifications, based on their emission lines. Specifically, sources are categorized into Seyfert 2 (Sy2), composite and LINER galaxies. Furthermore, we identify and select Compton-thick (CT) AGN using their X-ray and infrared luminosities at $12\mu m$. The data and the CT selection criteria are described in Sect. \ref{sec_data}. In Sect. \ref{sec_analysis}, we describe the process we follow to fit the spectral energy distributions (SEDs) of all the sources, using the CIGALE code, as well as the strict selection criteria we apply to select galaxies with robust SED fitting measurements. We perform a comprehensive investigation into SFR, M$_*$, accretion efficiency and stellar populations to decipher the intricate connections between AGN activity and the broader processes governing galaxy evolution in the local universe. The results appear in Sect. \ref{sec_results}. In Sect. \ref{sec_discussion}, we discuss our main findings and compare them with prior studies. Finally, Sect. \ref{sec_conclusions} presents a summary of our main findings. 

\begin{table*}
\caption{Models and the values for their free parameters used by X-CIGALE for the SED fitting of our galaxy sample. } 
\centering
\setlength{\tabcolsep}{1.mm}
\begin{tabular}{cc}
       \hline
Parameter &  Model/values \\
        \hline 
\multicolumn{2}{c}{Star formation history: delayed model with recent, constant burst/quench} \\
Age of the main population & 5000, 7000, 9000, 10000, 11000, 12000 Myr \\
e-folding time & 1000, 3000, 5000, 7000, 9000, 10000, 11000, 12000 Myr \\ 
Age of the burst/quench & 50, 100 Myr \\
Ratio of the SFR after/before the burst/quench episode & 0.01, 0.2, 0.8, 0.9, 1.0, 1.005, 1.015, 1.05, 1.10, 1.15, 1.20 \\
\hline
\multicolumn{2}{c}{Simple Stellar population: Bruzual \& Charlot (2003)} \\
Initial Mass Function & Salpeter\\
Metallicity & 0.02 (Solar) \\
\hline
\multicolumn{2}{c}{Galactic dust extinction} \\
Dust attenuation law & Charlot \& Fall (2000) \\
Reddening $A_V$ in the ISM & 0.001, 0.1, 0.2, 0.3, 0.4, 0.5, 0.6, 0.8, 1.0, 1.5 \\ 
\hline
\multicolumn{2}{c}{Galactic dust emission: Dale et al. (2014)} \\
$\alpha$ slope in $dM_{dust}\propto U^{-\alpha}dU$ & 2.0 \\
\hline
\multicolumn{2}{c}{AGN module: SKIRTOR} \\
Torus optical depth at 9.7 microns $\tau _{9.7}$ & 3, 7, 11 \\
Torus density radial parameter p ($\rho \propto r^{-p}e^{-q|cos(\theta)|}$) & 1.0 \\
Torus density angular parameter q ($\rho \propto r^{-p}e^{-q|cos(\theta)|}$) & 1.0 \\
Angle between the equatorial plan and edge of the torus & $40^{\circ}$ \\
Ratio of the maximum to minimum radii of the torus & 20 \\
Viewing angle  & 50\degree, 70\degree, 90\degree (type 2) \\
Accretion disk spectrum & Schartmann (2005) \\
AGN fraction & 0.0, 0.01, 0.1, 0.2, 0.3, 0.4, 0.5, 0.6, 0.7, 0.8, 0.9 \\
Extinction law of polar dust & SMC \\
$E(B-V)$ of polar dust & 0.0, 0.2, 0.4 \\
Temperature of polar dust (K) & 100 \\
Emissivity of polar dust & 1.6 \\
\hline
\label{table_cigale}
\end{tabular}
\tablefoot{For the definitions of the various parameters, see Sect. \ref{sec_sed}.}
\end{table*}
 
\section{Data}
\label{sec_data}

\subsection{Sample selection}
\label{sec_parent}
Our goal is to select a sample of low redshift, obscured AGN observed in X-rays while minimizing the contamination of star-forming galaxies. To this end we follow the criteria presented in \citet{Zhang2023} for selecting type 2 AGN in the SDSS: we selected all objects in the SDSS-DR18 \citep{Almeida2023} spectroscopic database (specObj table) that according to the SDSS pipeline have been classified as galaxies and with subclass AGN. We restrict our sample to galaxies with redshift below 0.1. In order to exclude local galaxies with a large extension where the SDSS photometry is highly unreliable, we imposed an additional redshift cut of $z > 0.02$ (with this limit the spectrograph fiber covers $\gtrsim1$~kpc of the nuclear region). We also selected only objects with good quality spectrum, i.e. with a median signal-to-noise ratio larger than 10 and no warnings in the estimated redshift. Finally, we kept only objects with a primary entry in the photometric database of SDSS and included in the Portsmouth catalogue of stellar kinematics and emission-line flux measurements \citep{Thomas2013}. This catalogue models the SDSS spectrum to derive the emission line properties and gives a reliable spectral classification (Seyfert, LINER, Composite, etc) based on BPT diagrams for each galaxy.

The final query that reflects this selection is:
\begin{verbatim}
SELECT * FROM specObj AS sp
JOIN PhotoObj AS ph 
 ON sp.bestObjID=ph.objID
JOIN emissionLinesPort AS ln 
 ON sp.specObjID=ln.specObjID 
WHERE
 sp.class='galaxy' AND sp.subclass='AGN'
 AND (sp.z BETWEEN 0.02 AND 0.1) 
 AND sp.zwarning=0 AND sp.snmedian>10
\end{verbatim}
This query returns a total of 7382 sources in the SDSS-DR18 database. In order to select sources observed in X-rays, we query the position of each galaxy in the RapidXMM database \citep{Ruiz2022}. This system provides X-ray flux upper-limits for all positions in the sky that have been observed by XMM-Newton. As of May 2023\footnote{XMM-Newton observations are ingested into the RapidXMM system when the data becomes public, so queries at a later date can give a larger number.}, we found 479 objects, out of our initial 7382 galaxies, in fields observed by XMM-Newton. Out of these 479 sources, 210 have a counterpart within 5~arcsec in the 4XMM-DR13 catalogue \citep{Webb2020}. The 4XMM-DR13 catalogue was built using XMM-Newton observations released up to 2022 December 31st. Our sample contains sources in 22  observations that were not included in the 4XMM-DR13. For these cases we used the X-ray source catalogues generated by the XMM-Newton pipeline, available in the archive, finding six additional sources with X-ray counterparts. In total, about 45 per cent (216 sources out of 479 sources) of our final sample is detected in X-rays.

\subsection{Photometry}
\label{sec_photometry}
In order to perform the SED analysis described in Sect.~\ref{sec_sed}, we need to build SEDs with good photometric coverage. In the ultra-violet (UV), we search for counterparts in the Revised Catalog for the GALEX All-Sky Imaging Survey \citep[GALEX-AIS,][]{Bianchi2017}. In the mid-infrared (MIR) regime, we used the AllWISE Catalog \cite{Cutri2013}. For the near-infrared (NIR) photometry, we relied on three catalogues. Most of our sources have counterparts in the 2MASS Extended Source Catalog \citep[2MASS-XSC,][]{Skrutskie2006}. When no counterpart was found in the 2MASS-XSC, we used the UKIDSS-DR11plus Large Area survey catalog \citep[UKIDSS-LAS,][]{Lawrence2007} if that sky region was covered by this survey, otherwise we search for a counterpart in the 2MASS Point Source Catalog \citep[2MASS-PSC,][]{Skrutskie2006}.

Given the low redshift of our selected sample, most of our sources are extended sources, clearly resolved in the optical and NIR bands. For a correct estimation of the galaxy properties we need to select measurements of the magnitudes that recover the emission of the whole galaxy. We used Petrosian magnitudes for the five SDSS bands (u, g, r, i, z), as recommended for photometry of nearby galaxies. We used the 'best' FUV and NUV GALEX magnitudes as recommended in the AIS catalog documentation. For sources with 2MASS-XSC photometry we used the isophotal J, H and K magnitudes. In the case of UKIDSS-LAS, we used the Petrosian magnitudes for the J, H and K bands. In AllWISE, we used the elliptical aperture magnitudes for the four WISE bands (W1, W2, W3, W4) when available, otherwise we used the profile-fitting magnitudes. We visually inspected the images in the different bands and the constructed SEDs to check that the different apertures used for measuring the magnitudes covered the same region of the galaxy.

\section{Analysis}
\label{sec_analysis}

\subsection{Galaxy properties}
\label{sec_sed}

To compute the properties of AGN and their host galaxies (e.g., AGN bolometric luminosity, SFR, M$_*$), we employ SED fitting through the CIGALE algorithm \citep{Boquien2019, Yang2020, Yang2022}. We adhere to the same templates and parametric grid in the SED fitting process as utilized in prior works \citep[e.g.][]{Koutoulidis2022, Mountrichas2022a}. In summary, the galaxy component is modeled using a delayed Star-Formation History (SFH) model with a functional form $\rm SFR\propto t \times exp(-t/\tau)$. A continuous star-formation period of 50 Myr is incorporated as a star-formation burst \citep{Malek2018, Buat2019}. Stellar emission is modeled using the single stellar population templates of \cite{Bruzual_Charlot2003}, attenuated in accordance with the \cite{Charlot_Fall_2000} attenuation law. To model nebular emission, CIGALE adopts nebular templates based on \cite{VillaVelez2021}. The emission from dust heated by stars is modeled following \cite{Dale2014}, excluding any AGN contribution. The AGN emission is included using the SKIRTOR models of \cite{Stalevski2012, Stalevski2016}. The parameter space employed in the SED fitting process is presented in Table~\ref{table_cigale}.

\begin{table}
\caption{Number of sources included within each AGN population considered in our analysis}
\centering
\setlength{\tabcolsep}{4mm}
\begin{tabular}{cc}
 \hline
{AGN population } & {number of sources} \\
 \hline
{Sy2 (no CT)} &  118 \\
{Composite} & 82 \\ 
{LINERS} &  124 \\
{CT} &  14\\
  \hline
\label{table_final_sources}
\end{tabular}
\end{table}

\begin{table*}
\caption{List of candidate Compton-thick sources}
\centering
\setlength{\tabcolsep}{2mm}
\begin{tabular}{ccccccc}
 \hline
  Name & $\alpha$ & $\delta$  & redshift & $\rm F_{2-10 keV}$ & $f_{AGN}$ & Reference \\
  (1)  & (2) & (3) & (4) & (5) & (6) & (7) \\
 \hline
NGC5765               &   222.71464   &  5.11449  &  0.0279 & $1.6\times 10^{-13}$ & 0.42 & \cite{Masini2019} \\
2XMMJ121742.0+034632  &184.42496 & 3.77529 & 0.0799 & $2.3\times10^{-16}$ & 0.75 &   \\
CGCG022-033          & 235.79429 & 1.32884 &  0.0397 & $4.9\times10^{-14}$ & 0.45 &  \\
IC 2227               &  121.77990 & 36.2330 & 0.0320  & $2.1\times10^{-13}$  & 0.33 & \cite{Silver2022} \\
2MASSJ132904.5+560353 & 202.26895 & 56.06481 & 0.043 & $2.4\times10^{-16}$ & 0.07 \\
SDSSJ230231.1+000147 &  345.62978  & 0.02990  & 0.095 & $3.6\times 10^{-16} $  & 0.60 &  \\
2MASSJ11435.6+153341 &  175.59826  &  15.5614 & 0.044 & $1.7\times10^{-15}$   & 0.28 &  \\
2MAXSJ141415.0+265812 & 213.56277 &  26.97002   &  0.066 &  $1.3\times10^{-14} $ & 0.47 & \\
2MASSJ145631.3+243635 & 224.13056  &  24.60972  &  0.033 & $2.5\times10^{-14}$  & 0.41 & \\
WISEA J120749.5+251236.2  &  181.95637   &  25.21005  &  0.098 & $2.4\times10^{-15}$   & 0.43 &  \\
LEDA1593164          &  238.95164  &  11.40939 &   0.072   &  $<5.8\times10^{-15}$ & 0.36 & \\
MCG-02-05-022       &   24.27895 &   -9.14931 &  0.070   &  $6.3\times10^{-15}$     &  0.63 &  \\
LEDA1373882          & 165.03945  &  10.05338   &   0.064   &   $8.6\times10^{-15}$ & 0.77 & \\
LEDA1169610          &  136.94548 &  0.57513     &  0.0536  &   $2.3\times10^{-15}$ &  0.18 & \\
  \hline
\label{CTcandidates}
\end{tabular}
\tablefoot{(1): Name (2, 3): optical right ascension and declination [degrees]. (4): Spectroscopic redshift. (5): X-ray flux in the 2-10 keV band  ($\rm erg~s^{-1}$).  (6): AGN fraction according to the CIGALE spectral energy distribution fit. 
(7): Reference}
\end{table*}

\begin{figure*}
\centering
  \includegraphics[width=0.85\textwidth, height=14cm]{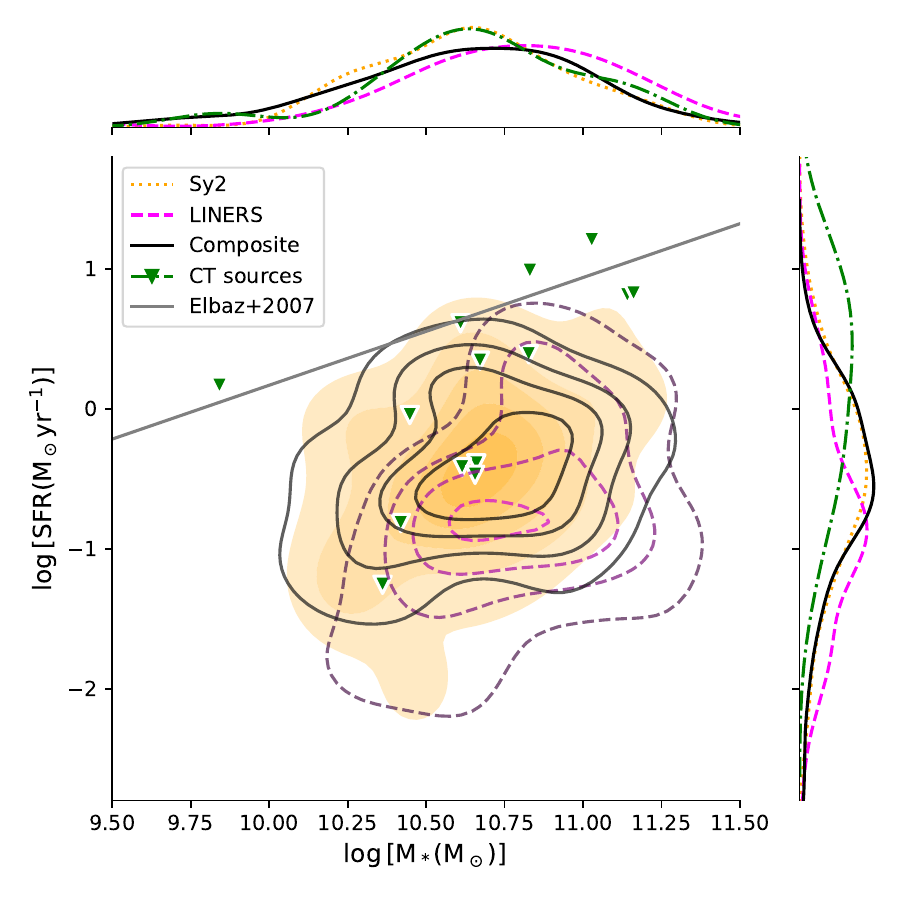}   
  \caption{Distribution of sources in the SFR-M$_*$ plane. Different AGN populations are presented with different colours and lines, as indicated in the legend of the plot. The solid, grey line, shows the local SFR-M$_*$ relation presented in \cite{Elbaz2007}, for SDSS galaxies.}
  \label{fig_sfr_mstar}
\end{figure*}  

\begin{figure}
\centering
  \includegraphics[width=1.\columnwidth, height=7.2cm]{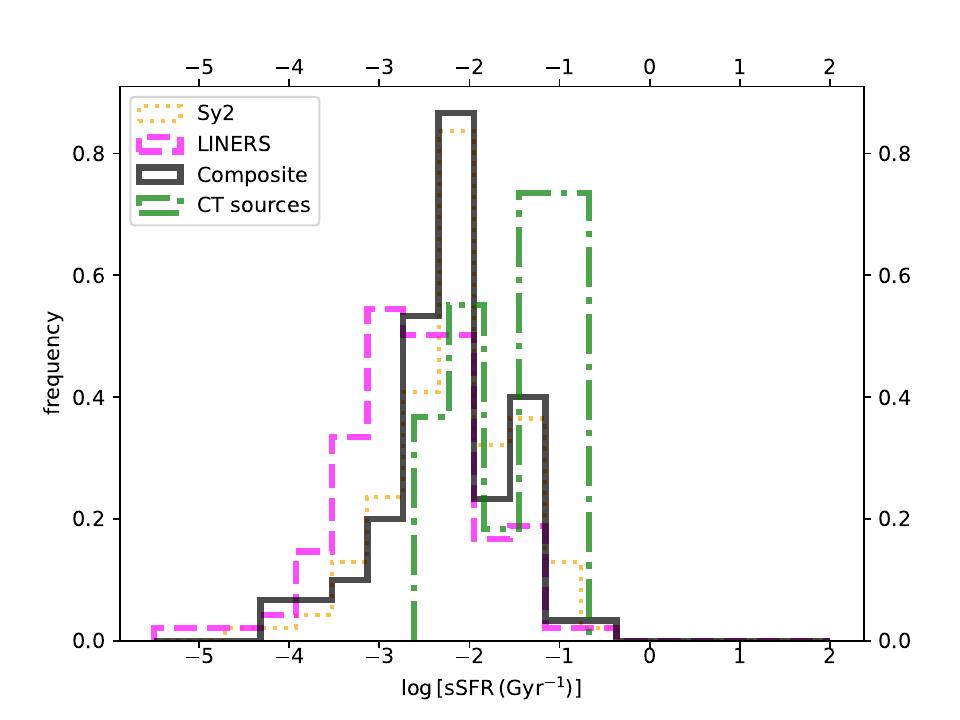} 
  \caption{The distribution of sSFR ($\rm \frac{SFR}{M_*}$) for the different AGN populations, used in our study.}
  \label{fig_ssfr}
\end{figure}  

\begin{figure*}
\centering
  \includegraphics[width=0.85\textwidth, height=12cm]{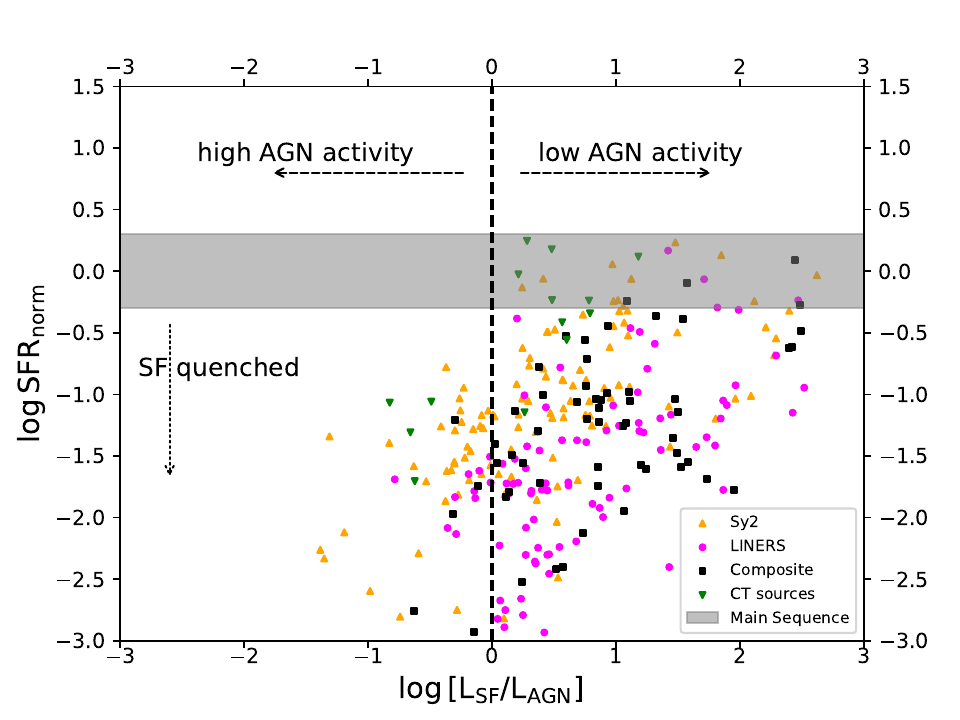} 
  \caption{SFR$_{norm}$ as a function of the ratio of the star-formation luminosity (L$\rm _{SF}$) to the AGN luminosity (L$\rm _{AGN}$). SFR$_{norm}$ is defined as the ratio of the SFR of the AGN to the SFR calculated using the \cite{Elbaz2007} expression for SDSS galaxies. The grey shaded area indicates an area $\pm 0.3$\,dex around $\rm log\,SFR_{norm}=0$ to denote the main-sequence. Below the grey area, the star-formation of the AGN host galaxies is quenched. The dashed vertical lines indicates the $\rm log\frac{L_{SF}}{L_{AGN}}=0$. At $\rm log\frac{L_{SF}}{L_{AGN}}<0$ the AGN activity is the dominant mechanism in the host galaxy, whereas at $\rm log\frac{L_{SF}}{L_{AGN}}>0$ the AGN activity is low.}
  \label{fig_sfrnorm}
\end{figure*}  

\begin{figure}
\centering
  \includegraphics[width=1.\columnwidth, height=7.2cm]{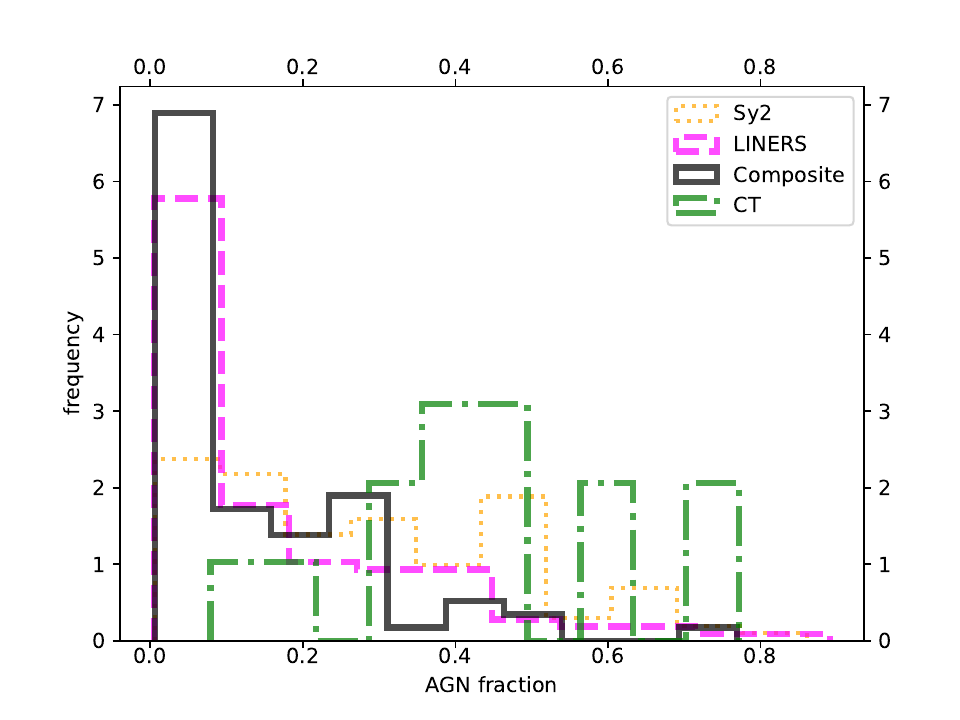} 
  \caption{Distributions of the AGN fractions, calculated by CIGALE, for the different AGN populations, as indicated in the legend. The AGN fraction is defined as the fraction of the total infrared emission coming from the AGN}
  \label{fig_fracAGN}
\end{figure}

\subsection{Reliability criteria}
\label{sect_criteria}

To ensure the reliability of our analysis, we implement selection criteria akin to those employed in prior studies \citep[e.g.,][]{Mountrichas2022b, Mountrichas2022c, Buat2021, Pouliasis2022, Mountrichas2023d}. Specifically, in order to exclude sources with unreliable SED fitting measurements and host galaxy information, we set a reduced $\chi ^2$  threshold of $\chi ^2_r <5$ \citep[e.g.,][]{Masoura2018, Buat2021}. This criterion led to the exclusion of six sources from our dataset. Additionally, we omit systems for which the CIGALE algorithm could not constrain the SFR and M$_*$. For this purpose, we leverage the two values provided by CIGALE for each estimated galaxy property. One value corresponds to the best model, while the other (bayes) represents the likelihood-weighted mean value. A substantial disparity between these two calculations implies a complex likelihood distribution and significant uncertainties. Consequently, we only incorporate sources in our analysis that satisfy the conditions $\rm \frac{1}{5}\leq \frac{SFR_{best}}{SFR_{bayes}} \leq 5$ and $\rm \frac{1}{5}\leq \frac{M_{*, best}}{M_{*, bayes}} \leq 5$, where SFR$\rm _{best}$ and  M$\rm _{*, best}$ are the best-fit values of SFR and M$_*$, respectively and SFR$\rm _{bayes}$ and M$\rm _{*, bayes}$ are the Bayesian values estimated by CIGALE. 

There are 338 sources that meet the specified criteria. Among these sources we identify and select 14 CT AGN candidates (see next section). From the remaining 324 galaxies, 118 are classified as Sy2, 82 as composite and 124 as LINER galaxies, based on the Portsmouth catalogue \citep{Thomas2013}. These are the sources used in our analysis (Table \ref{table_final_sources}).

\begin{figure}
\centering
 \includegraphics[width=1\columnwidth, height=7.2cm]{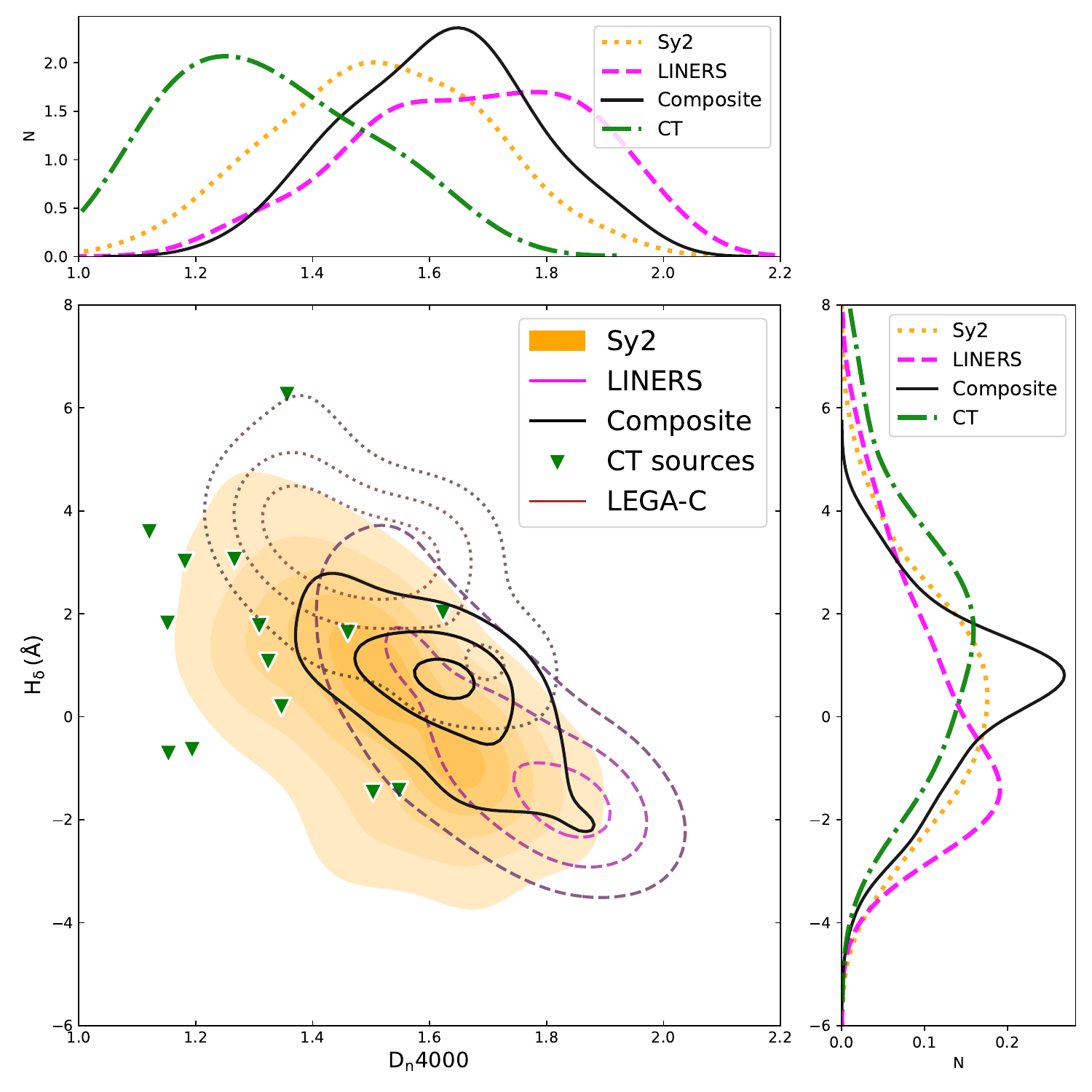} 
  \caption{Distribution of the different AGN populations in the H$_\delta-$D$_n$4000 space. We also plot the results for the heavily obscured ($N_\mathrm{H} > 10^{23}~\mathrm{cm^{-2}}$) AGN sample used in \cite{Georgantopoulos2023}, at $\rm z\sim 1$ (LEGA-C, brown, dotted contours).}
  \label{fig_hd_d4000}
\end{figure}  

%\begin{figure}
%\centering
% \includegraphics[width=1.\columnwidth, height=9cm]{halpha_d4000_all_classes_CT_with_contours_weights_LEGAC_v2.pdf} 
%  \caption{Distribution of the different AGN populations used in our analysis, in the H$_\delta-$D$_n$4000 space, compared to the heavily obscured ($N_\mathrm{H} > 10^{23}~\mathrm{cm^{-2}}$) AGN sample used in \cite{Georgantopoulos2023}, at $\rm z\sim 1$.}
%  \label{fig_hd_d4000_legac}
%\end{figure}  

\subsection{Selection of Compton-Thick AGN candidates}
\label{sect_ct}
One of the goals of this work is the identification of potential CT AGNs and studying the properties of their host galaxy in comparison with the overall properties of the type 2 AGN population. CT AGN  show X-ray absorption with Hydrogen column densities $N_\mathrm{H} > 10^{24}~\mathrm{cm^{-2}}$ that largely suppress the direct X-ray emission below 10~keV
 \citep{Ricci2015, Akylas2015, Georgantopoulos2019, Torres-Alba2021}.

In order to identify CT candidates we follow the work of \citet{Pfeifle2022}, where they presented a diagnostic for the X-ray absorption in AGN, based on the ratio of the mid-infrared and the $\rm 2-10~keV$ X-ray luminosities.
As the mid-infrared luminosity represents a reliable proxy of the isotropic AGN emission, a low X-ray to mid-infrared 
luminosity ratio provides a powerful method to identify CT sources \citep{Alexander2008,Georgantopoulos2011,Rovilos2014}. 
\cite{Pfeifle2022} use the BAT AGN Spectroscopic Survey \citep[BASS][]{Koss2017,Koss2022,Ricci2017,Ichikawa2017} which includes sources detected in the ultrahard X-ray  band (14–195~keV) and it is expected to be a complete census, independent of X-ray obscuration, of the most luminous AGN in the local Universe.

Using the same data set and methods of \citet{Pfeifle2022}, we estimated the expected completeness and purity levels for CT samples using different X-ray-to-12-$\mu$m luminosity ratios, as shown in Fig.~\ref{fig_ct_diagnostic}. Our results show that for $\log(L_\mathrm{obs}(2-10~\mathrm{keV}) / L(12~\mu\mathrm{m})) < -1.6$ we can expect a $\sim80$ per cent completeness with a purity of $\sim80$ per cent.

We estimated the X-ray-to-12-$\mu$m luminosity ratios for our sample of low-z type 2 AGNs. Since this is a spectroscopically selected sample of nearby objects, it includes objects with low luminosity AGNs and the host galaxy emission dominates even in the MIR range. Hence, in order to avoid contamination due to the host galaxy, the 12~$\mu$m luminosity we used for estimating the \citet{Pfeifle2022} diagnostic is the corresponding to the AGN emission we obtained in our SED analysis using CIGALE (see Sect.~\ref{sec_sed}). 

%Figure~\label{fig_ct_selection} shows The evidence indicates that the different types of AGN we've examined may result from distinct phases of AGN activity. For example, if a supermassive black hole (SMBH) becomes active early in a galaxy's evolution, when there's plenty of gas, it might exhibit a high accretion rate, appearing as a Seyfert 2 (Sy2) and potentially transitioning from Compton-thick (CT) to Sy2. On the other hand, if the AGN activity begins later in the galaxy's timeline when gas is less abundant, the SMBH would likely have a lower accretion rate, leading to a Low-Ionization Nuclear Emission-line Region (LINER).

Accordingly to the \citet{Pfeifle2022} diagnostic we discussed above, we considered CT candidates those sources below the $\log(L_\mathrm{obs}(2-10~\mathrm{keV}) / L(12~\mu\mathrm{m})) = -1.6$ line and optically classified as Sy2 galaxies (see Fig.~\ref{fig_ct_selection}). We found a total of 14 CT candidates.
Our CT sample is by no means complete. 
This is because we have discarded sources with low quality optical spectra i.e. signal to noise ratio lower than 10. Some 
 CT sources especially the fainter ones may be among the discarded sources. In addition, our selected $\log(L_\mathrm{obs}(2-10~\mathrm{keV}) / L(12~\mu\mathrm{m})) < -1.6$ criterion 
 can find only 80\% of the known CT AGN while among the selected CT AGN only 80\%
 are bona-fide CT AGN. Finally, the low $\log(L_\mathrm{obs}(2-10~\mathrm{keV}) / L(12~\mu\mathrm{m})) $ ratio criterion may be sensitive to other types of sources such as turnoff AGN, see for example the discussion in \cite{Saade2022}. 
 
 Next we check in the literature whether  our sources that have reasonably good quality X-ray observations available are associated indeed with CT AGN. Out of  our 14 sources, three have been observed by {\it NuSTAR}: NGC 5765, IC2227 and LEDA 1373882. \cite{Masini2019} find that NGC 5765 is a reflection dominated CT AGN with $\rm N_H\sim10^{25}\,cm^{-2} $. The {\it NuSTAR} observations of IC 2227 have been reported by \citet{Silver2022}. They find that the source is heavily obscured with $\rm N_H\sim 3\times10^{23} cm^{-2}$. The remaining source has not been detected by {\it NuSTAR}. Next, we search whether there is additional information in the literature 
 regarding the X-ray spectra of the remaining 11 sources. The vast majority of these are faint sources, having fluxes below $5\times 10^{-14}$ $\rm erg~cm^{-2}s^{-1}$ thus impeding the extraction of good quality spectra. 
%  The {\it XMM-Newton} and {\it Chandra} spectra of IC 750 have been presented in \cite{Zaw2020}. 
%  The {\it Chandra} image shows a nuclear source and three more sources close to the nucleus. 
%  These sources are confused in the {\it XMM-Newton} image owing to its limited spatial resolution. 
%  \cite{Zaw2020} find that the {\it Chandra} spectrum of the nuclear source is absorbed by a small column density
%  of $\rm N_H\sim 10^{21} ~cm^{-2}$, but see \cite{Chen2017}  who find a column density about two orders of magnitude higher.
%  Despite the uncertain X-ray column density, \cite{Zaw2020} point out that
%  IC 750 may hide a  deeply buried Compton-thick nucleus as  it is associated with a water maser. Water maser systems are known to host  Compton-thick  AGNs \citep{Castangia2013}. 
%   The {\it XMM-Newton} spectrum of NGC3185 is reported in \cite{Cappi2006}. 
%   The X-ray spectrum is consistent with an unabsorbed source but the $F_{2-10keV}/F[OIII]$ diagnostic suggests 
%   that the source is Compton-thick. 
One of our candidate sources (LEDA1593164) has not been detected and there is only an upper limit in X-ray flux available \citep[see e.g.,][]{Ruiz2021}. Three of our sources  are associated with targets: NGC5765, IC2227, and 2MASSX1390454+5603528. 
% Finally, \cite{Akylas2024} fitted the X-ray spectrum of CGCG 074-129 and found a column density of $\rm 6\times 10^{23}\,cm^{-2}$ and intrinsic $\rm log\,[L_{X,2-10keV}(ergs^{-1})] = 41.9$. 
In Table \ref{CTcandidates} we give the full list of our CT candidate sources.

\section{Results}
\label{sec_results}

In this section, we explore the location of diverse AGN populations in relation to the star-forming main-sequence (MS) and delve into the influence of SMBH activity on this location. Additionally, we conduct a comparison of their stellar populations and we analyze the accretion power exhibited by our sources.

\begin{table*}
\caption{Median values and their 25th and 75th percentiles for the SFR, M$_*$, sSFR, D$_n$4000, H$_\delta$ and $\lambda _{sBHAR}$ of the different AGN populations examined in our study.}
\centering
\setlength{\tabcolsep}{0.3mm}
\begin{tabular}{ccccccc}
 \hline
{AGN population } & $\rm log\,[SFR (M_\odot yr^{-1})]$ & $\rm log\,[M_*(M_\odot)]$  & $\rm log\,sSFR(Gyr^{-1})$ & D$_n$4000 & H$_\delta$ & log\,$\lambda _{sBHAR}$\\
 \hline
{Sy2} & -0.49\,[-0.92, -0.03] & 10.64\,[10.41, 10.84] & -2.15\,[-2.52, -1.75]  & 1.51\,[1.39, 1.66] & 0.35\,[-1.15, 1.71] & -2.57\,[-3.08, -2.26]\\
{composite} & -0.54\,[-0.85, -0.08] & 10.62\,[10.42, 10.90] & -2.13\,[-2.58, -1.92] & 1.62\,[1.50, 1.73] & 0.50\,[-0.77, 1.15] & -3.21\,[-3.61, -2.78] \\ 
{LINERS} & -0.87\,[-1.32, -0.36]  & 10.81\,[10.56, 11.02] & -2.59\,[-3.07, -2.15] &  1.71\,[1.53, 1.84] & -0.35\,[-1.65, 1.45] &  -3.28\,[-3.69, -2.80]\\
{CT} & 0.25\,[-0.41, 0.76]  & 10.66\,[10.49, 10.83] & -1.39\,[-2.04, -1.08] & 1.32\,[1.18, 1.43] & 1.69\,[-0.44, 2.76] & -1.82\,[-2.10, -1.60]\\
  \hline
\label{table_sfr_mstar}
\end{tabular}
\end{table*}

\begin{figure}
\centering
 \includegraphics[width=1.1\columnwidth, height=7.5cm]{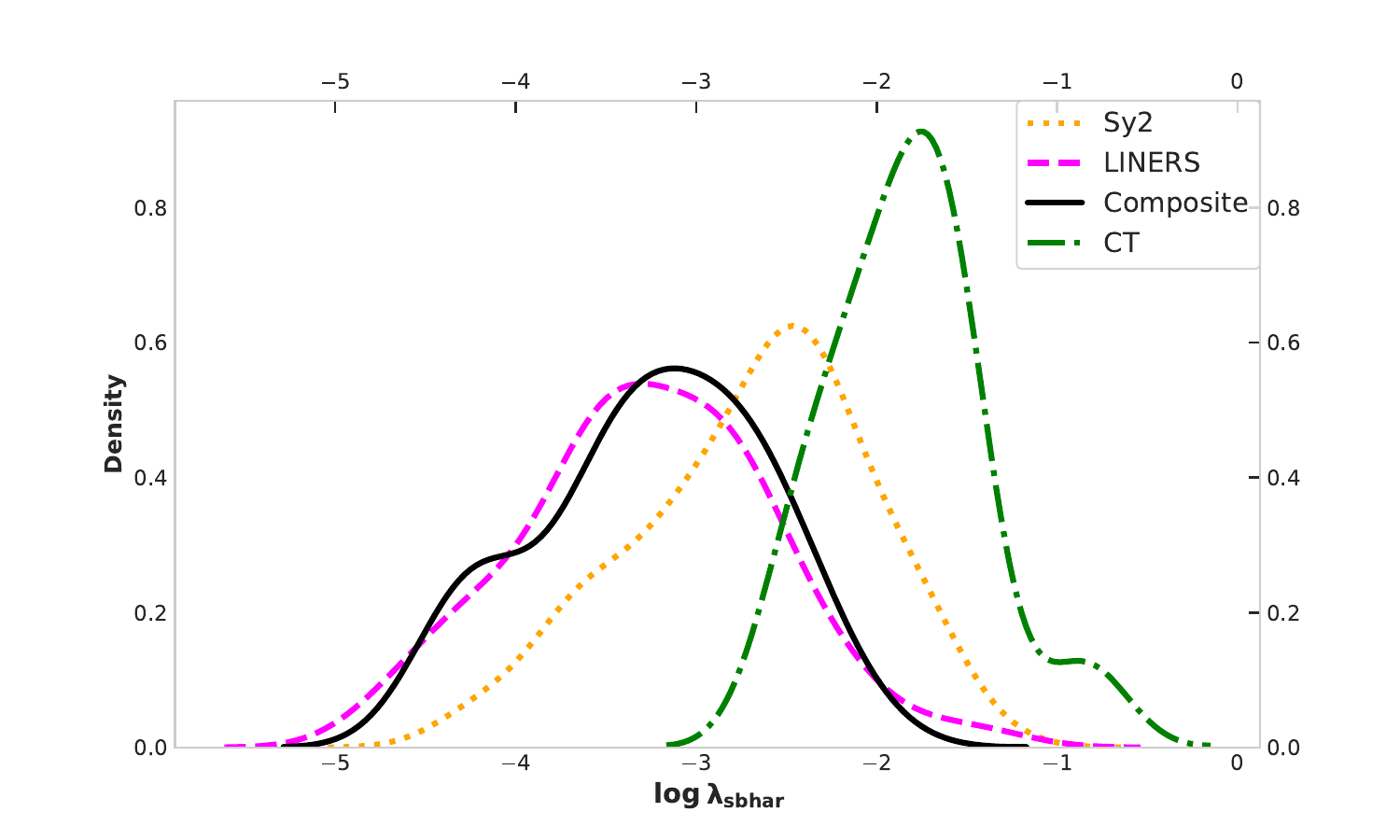} 
 \includegraphics[width=1.1\columnwidth, height=9cm]{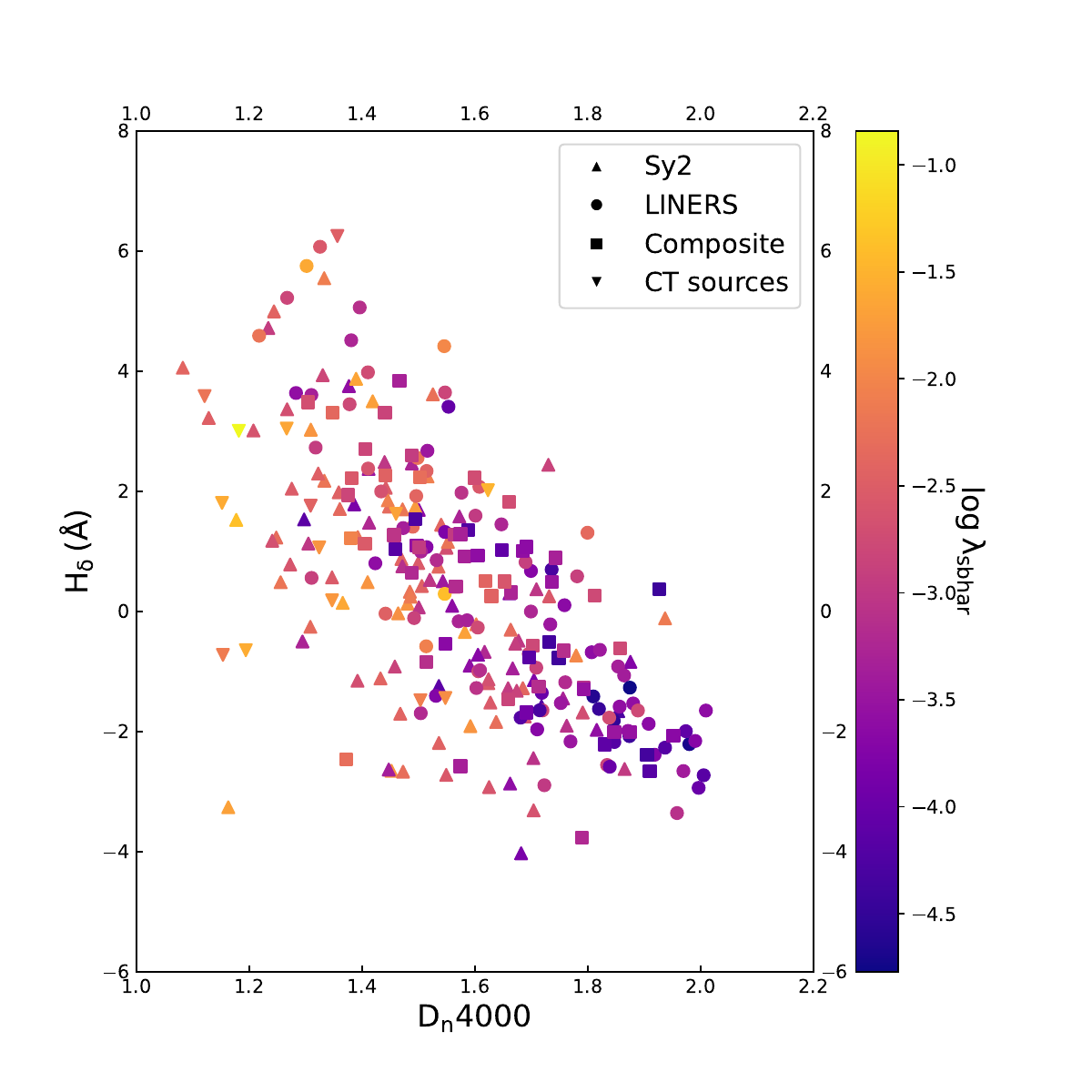} 
  \caption{The specific black hole accretion rate, $\lambda _{sBHAR}$, for the different AGN populations. The top panel shows the distribution of $\lambda _{sBHAR}$ for the different AGN classes. The bottom panel shows the distribution of our sources in the H${_\delta}-$D$_n$4000 space, colour-coded based on their $\lambda _{sBHAR}$ values.}
  \label{fig_lambda}
\end{figure}  

\subsection{The position of the AGN classes relative to the main-sequence}
\label{sec_ms}

To examine the relative positioning of the different AGN populations in relation to the star-forming MS, we first investigate the distribution of our selected galaxies in the SFR-M$_*$ plane (Sect. \ref{sec_sfr_mstar}) and then we compare the SFR of the sources in our dataset to the SFR of star-forming MS galaxies, as a function of luminosity (Sect. \ref{sec_sfrnorm_vs_lum}).

\subsubsection{Distribution of sources in the SFR-M$_*$ plane}
\label{sec_sfr_mstar}

In Fig. \ref{fig_sfr_mstar}, we illustrate the distribution of the AGN populations in the SFR$-$M$_*$ plane. Additionally, we incorporate the local SFR$-$M$_*$ relation, as determined from SDSS galaxies by \cite{Elbaz2007}, represented by the grey line for reference. Notably, the majority of our sources appear below this line, indicating that our sources predominantly inhabit quiescent systems. Table \ref{table_sfr_mstar} provides median values and their corresponding 25th and 75th percentiles for each host galaxy property and AGN class. Intriguingly, LINERS galaxies exhibit the highest M$_*$ (by $\sim 0.2$\,dex) and the lowest SFR compared to other AGN classes. Among the four AGN classes, CT sources display the highest median SFR values. Despite these disparities, we note that Kolmogorov-Smirnov tests (KS-tests) indicate that these differences lack statistical significance (i.e., $<2\,\sigma$), as the $\rm p-values$ obtained range from $0.2-0.9$ (where a $\rm p-value=0.05$ signifies a statistical significance of $\sim 2\,\sigma$). Similar outcomes are observed with other statistical tests, such as Mann-Whitney, Anderson, and Kuiper tests.

Fig. \ref{fig_ssfr} depicts the distributions of sSFR ($\frac{SFR}{M_*}$) for different AGN classes, with median values and their corresponding 25th and 75th percentiles provided in Table \ref{table_sfr_mstar}. LINER galaxies exhibit the lowest sSFR compared to other AGN populations, which display comparable median values and sSFR distributions, with the exception of CT AGN that display the highest median sSFR values. Notably, despite the $\rm p-values$ obtained from the comparison of LINERS' sSFR distribution with other AGN populations being relatively lower (ranging from $0.1-0.2$) compared to those for SFR and M$_*$ distributions, these differences do not achieve statistical significance at a $2\,\sigma$ level.

\subsubsection{SFR$_{norm}$ vs. luminosity}
\label{sec_sfrnorm_vs_lum}

An alternative way to illustrate the position of AGN relative to the MS, is to calculate the SFR$_{norm}$ parameter \citep[e.g.,][]{Mullaney2015, Masoura2018, Masoura2021, Bernhard2019, Koutoulidis2022, Pouliasis2022, Mountrichas2022b, Mountrichas2023c, Mountrichas2023d}. SFR$_{norm}$ is defined as the ratio of the SFR of AGN to the SFR of star-forming MS galaxies, with comparable M$_*$ and redshift. Therefore, SFR$_{norm}>1$ indicates that the AGN is located above the MS, whereas SFR$_{norm}<1$ indicate that the AGN is below the MS. For the calculation of SFR$_{norm}$, we utilize the expression derived in \cite{Elbaz2007}, that used SDSS galaxies in the local Universe. It is important to note that using analytical expressions from existing literature for estimating SFR$_{norm}$ may introduce systematic biases, as opposed to employing galaxy control samples \citep{Mountrichas2021c}. However, for the purposes of our analysis, these potential systematics do not impact our results and conclusions.

Fig. \ref{fig_sfrnorm} presents the distribution of our sources in the SFR$\rm _{norm}-\frac{L_{SF}}{L_{AGN}}$ space. $\rm L_{SF}$ and $\rm L_{AGN}$ are the luminosities originating from the star-formation and the AGN, respectively. Both parameters are defined as the integrated luminosities in the range between 8 and 1000 $\mu$m. To gauge the accuracy of CIGALE's estimations for these parameters, we can check how well CIGALE calculates the AGN fraction, $\rm frac_{AGN}$. This is because $\rm frac_{AGN}$ is defined as the fraction of the total infrared emission coming from the AGN and therefore is derived from data within similar wavelengths as these two parameters. In Fig. \ref{fig_fracAGN}, we present the distributions of $frac_{AGN}$ for the four AGN populations. This plot demonstrates the considerable range of AGN activity present in our sources. Sy2 and CT appear to have the highest $frac_{AGN}$ values (median values of 0.28 and 0.44, respectively) compared to composite and LINER galaxies (median values of 0.14 and 0.17, respectively).

To evaluate the accuracy of $\rm frac_{AGN}$, we use mock catalogues generated by CIGALE based on the best-fitting model for each source in our dataset. CIGALE essentially creates a mock sample by taking the best-fitting flux for each source and introducing noise to it, which is derived from a Gaussian distribution with the same standard deviation as the observed flux. The mock data are then analyzed in the same manner as the actual observations. The precision of each estimated parameter can be assessed by comparing the original input values to the output values from the analysis (ground truth versus estimated value).

Our investigation revealed that the difference between the original and estimated values of the AGN fractions has a mean value of 0.05 (median value of 0.02), with a dispersion of 0.16. When we focus on sources with low AGN fraction values (less than 0.2), the mean difference is 0.04 (median difference is 0.03), with a dispersion of 0.08. Given these findings, we consider the calculated AGN fractions, and by extension the luminosities for star formation ($L_{SF}$) and active galactic nuclei ($L_{AGN}$), to be reliable. 

Since we have addressed the reliability of the $L_{SF}$ and $L_{AGN}$ parameters, we now examine the distribution of our sources in the  SFR$\rm _{norm}-\frac{L_{SF}}{L_{AGN}}$ space (Fig. \ref{fig_sfrnorm}). In systems that have $\rm \frac{L_{SF}}{L_{AGN}}<1$ (or $<0$ in logarithmic space) the AGN activity dominates over star-formation activity, while, $\rm \frac{L_{SF}}{L_{AGN}}>1$  implies either a more starburst-dominated galaxy or relatively low AGN luminosities \citep{Netzer2009}. The grey shaded area indicates a $\pm 0.3$\,dex around the MS (i.e., around SFR$_{norm}=1$). We notice that regardless of the AGN class, in all cases SFR$_{norm}$ increases with  $\rm \frac{L_{SF}}{L_{AGN}}$, signifying that the AGN host galaxy is closer to the MS for higher values of $\rm \frac{L_{SF}}{L_{AGN}}$. This trend could stem from either an elevation in $\rm L_{SF}$, a reduction in $\rm L_{AGN}$, or a combination of both factors as galaxies approach the MS. To discern the primary contributor of what drives the AGN towards or away from the MS, we compute the slopes of the SFR-M$_*$ relation for various AGN populations (see Fig. \ref{fig_sfr_mstar}) using the linmix module \citep{Kelly2007}. Linmix conducts a linear regression between two parameters by iteratively adjusting the data points within their uncertainties. The findings indicate that the slope gradually flattens as we transition from CT (slope of 1.24), to Sy2 (0.78), to composite (0.59), and ultimately to LINER galaxies (0.51). Therefore, AGN populations with higher AGN activity (based on their AGN fraction measurements, i.e., CT and Sy2) appear to have steeper slopes compared to AGN systems with lower AGN activity (i.e., composite and LINER galaxies).  This may suggest that the AGN activity is aiding in quenching the SFR in the examined systems.

%Given the low levels of star-formation activity in our datasets, this increase in the $\rm \frac{L_{SF}}{L_{AGN}}$ ratio primarily results from a decrease in AGN luminosities. 

%These findings suggest that lower AGN activity (i.e., higher $\rm \frac{L_{SF}}{L_{AGN}}$ ratio) positions the host galaxy closer to the MS (i.e., SFR$_{norm}\sim 1$). Additionally, it is observed that most composite and LINER galaxies are located to the right of Sy2 and CT sources, attributable to the larger AGN fractions exhibited by Sy2 and CT galaxies (median values of 0.28 and 0.44, respectively) compared to composite and LINER galaxies (median values of 0.14 and 0.17, respectively). 

%We also note that by confining the CT subset to encompass only CT Sy2 sources, reduces the number of sources with $\rm frac_{AGN}<0.1$ from six to one.

Overall,  our analysis indicates that the majority of the sources examined in this study are positioned below the MS. LINERS preferentially inhabit galaxies characterized by higher stellar mass and lower levels of SFR activity compared to Sy2, composite, and CT sources. However, these distinctions do not reach statistical significance exceeding 2\,$\sigma$. We also find indications that CT sources may present enhanced levels of star-formation compared to non-CT AGN. Additionally, our results suggest that a lower level of AGN activity corresponds to a closer positioning of the host galaxy to the MS.

%\begin{table}
%\caption{Median values and their 25th and 75th percentiles for the D$_n$4000 and H$_\delta$ of the different AGN populations examined in our study.}
%\centering
%\setlength{\tabcolsep}{3mm}
%\begin{tabular}{ccc}
% \hline
%{AGN population } & D$_n$4000 & H$_\delta$\\
% \hline
%{Sy2} & 1.51\,[1.39, 1.66] & 0.35\,[-1.15, 1.71]\\
%{composite} & 1.62\,[1.50, 1.73] & 0.50\,[-0.77, 1.15] \\ 
%{LINERS} &  1.71\,[1.53, 1.84] & -0.35\,[-1.65, 1.45]\\
%{CT} & 1.32\,[1.18, 1.43] & 1.69\,[-0.44, 2.76] \\
%  \hline
%\label{table_sfr_mstar}
%\end{tabular}
%\end{table}

%\begin{table}
%\caption{Median values and their 25th and 75th percentiles for the $\lambda _{sBHAR}$ of the different AGN populations examined in our study.}
%\centering
%\setlength{\tabcolsep}{3mm}
%\begin{tabular}{ccc}
% \hline
%{AGN population } & log\,$\lambda _{sBHAR}$ \\
% \hline
%{Sy2} & -2.57\,[-3.08, -2.26] \\
%{composite} & -3.21\,[-3.61, -2.78]  \\ 
%{LINERS} &  -3.28\,[-3.69, -2.80] \\
%{CT} & -1.82\,[-2.10, -1.60]  \\
%  \hline
%\label{table_sfr_mstar}
%\end{tabular}
%\end{table}

\subsection{The stellar populations of the different AGN classes}
\label{sec_stellar}

Next, we conduct a comparative analysis of the stellar populations among galaxies hosting different AGN classes. In Fig. \ref{fig_hd_d4000}, we illustrate the distributions of the various AGN populations in the H$_\delta-$D$_n$4000 space. Recognizing that more massive systems tend to harbor older stars, we apply weights to these distributions based on the M$_*$ of the sources. Specifically, we assign a weight to each AGN to match the M$_*$ of the four AGN classes \citep[e.g.][]{Mountrichas2022c}. It is important to highlight that, while the D$_n$4000 measurements exhibit relatively small uncertainties (with a median uncertainty value representing approximately 10\% of the measured value), the uncertainties associated with H$_\delta$ are notably larger (with a median value of H$_\delta$ uncertainties being around 80\% of the measured value). Consequently, while we provide the distributions of the H$_\delta$ spectral line, our primary conclusions are derived from the results based on the D$_n$4000 spectral index due to its comparatively smaller errors.

Our findings indicate that LINER galaxies exhibit, on average, the oldest stellar populations compared to the other AGN classes. Sy2 and composite galaxies display stars of similar age, while CT sources showcase the youngest stars among the various AGN classes examined in our study. Although the statistical tests do not reveal significant differences based on the calculated $\rm p-values$, likely due to the broad distributions, the observed patterns in the distributions are notably distinct. The median values along with their corresponding 25th and 75th percentiles for each galaxy population are presented in Table \ref{table_sfr_mstar}. 

%We note that when confining the CT subset to encompass solely CT Sy2 galaxies, the median value of D$_n$4000 becomes even lower at 1.31, as opposed to the value of 1.38 reported in Table \ref{table_sfr_mstar} for the overall CT subset.

These findings align with the outcomes presented in the previous section. Specifically, LINER galaxies, characterized by the lowest star-formation activity among the various AGN classes, demonstrate the highest D$_n$4000 values, indicating they harbor the oldest stellar populations. Composite and Sy2 galaxies, which share similar levels of star-formation activity, also tend to possess comparable stellar populations. Furthermore, our current results are consistent with those in the prior section, underscoring that CT sources, on average, display heightened star-formation activity and host the youngest stellar populations among the AGN classes examined in this study.

In Fig. \ref{fig_hd_d4000}, we also juxtapose the distribution of our sources in the H$_{\delta}-$D$_n$4000 space with that of the heavily obscured LEGA-C AGN, as presented in \cite{Georgantopoulos2023} (illustrated by brown, dotted contours). In the study by \cite{Georgantopoulos2023}, 73 AGN in the COSMOS field were examined, with available measurements for their spectral indices obtained from the LEGA-C catalogue \citep{Wel2021} at redshifts within $\rm 0.6<z<1$. The investigation involved a comparison of various properties, including M$_*$, sSFR, Eddington ratio, and stellar populations, between heavily obscured and non-obscured AGN, using X-ray criteria for the classification of the sources and applying a threshold at $N_\mathrm{H} = 10^{23}~\mathrm{cm^{-2}}$. Notably, the LEGA-C AGN exhibit lower D$_n$4000 values (and higher H${_\delta}$ values) compared to our sample. It is crucial to acknowledge that the two AGN populations differ not only in terms of their redshifts but also in their classification criteria, their M$_*$ properties (see bottom panel of Fig. \ref{fig_ssfr}) and L$_X$ (LEGA-C AGN are about two orders of magnitude more luminous compared to the sources used in our analysis; see also the discussion in Sect. \ref{sec_discussion}).

\subsection{The accretion efficiency of different AGN classes}
\label{sec_eddington}

In this section, we investigate the accretion efficiency  across various AGN classes within our datasets. This efficiency is measured through the n$_{Edd}$. In cases where the M$_{BH}$ measurements are not available, the specific black hole accretion rate, $\lambda _{sBHAR}$, is used as a proxy of the n$_{Edd}$ \citep[e.g.,][]{Aird2018, Mountrichas2021c, Mountrichas2022a}.  For the calculation of  $\lambda _{sBHAR}$ the following expression is used:

\begin{equation}
\lambda_{sBHAR}=\rm \frac{L_{bol}}{1.26\times10^{38}\,erg\,s^{-1}\times0.002\frac{M_{*}}{M_\odot}}.  
\label{eqn_lambda}
\end{equation}
To calculate $\lambda _{sBHAR}$, we employ the measurements of $L_{bol}$ and M$_*$ provided by CIGALE. It is important to acknowledge that the effectiveness of $\lambda _{sBHAR}$ as a proxy for n$_{Edd}$ hinges on factors such as the scatter in the M$_{BH}-$M$_*$ relation and the accuracy of AGN bolometric luminosity estimates, as previous studies have shown \citep{Lopez2023, Mountrichas2023d}. Nonetheless, in our examination, we emphasize the comparison of $\lambda _{sBHAR}$ across distinct AGN classes rather than its absolute values.

Figure \ref{fig_lambda} displays the distributions of $\lambda _{sBHAR}$ for various AGN populations in our dataset (top panel). The corresponding median values and percentiles are shown in Table \ref{table_sfr_mstar}. Our findings indicate that LINER and composite galaxies showcase analogous $\lambda _{sBHAR}$ distributions and median values, which are also the lowest among the AGN classes considered. Sy2 galaxies tend to exhibit higher $\lambda _{sBHAR}$ values, while CT sources present the highest $\lambda _{sBHAR}$ values within the AGN populations in our sample. Utilizing the KS-test indicates that these distinctions hold statistical significance at a level exceeding $>2\,\sigma$, as the $\rm p-values$ range from $10^{-5}$ to $10^{-7}$. Comparable $\rm p-values$ are obtained through other statistical tests, including Mann-Whitney, Anderson, and Kuiper. 

%We note that restricting the CT subset to include only CT Sy2 galaxies, results in even higher $\lambda _{sBHAR}$ values (-1.81) compared to the total CT subset.

The bottom panel of Figure \ref{fig_lambda}, presents the distribution of the different AGN classes in the H$_\delta$-D$_n$4000 space, colour-coded based on the $\lambda _{sBHAR}$ of the sources. The results indicate that sources with younger stellar populations (i.e., D$_n4000<1.4$) tend to exhibit higher $\lambda _{sBHAR}$ values compared to sources with older stars. This is in line with the findings of \citet[][see their Fig. 4]{Georgantopoulos2023}.

\section{Discussion}
\label{sec_discussion}

In this work, we have identified CT sources and have investigated their properties as a different AGN class and compare it with the other AGN populations. We found indications that CT sources may present enhanced levels of star-formation activity, but, most importantly, our analysis revealed that CT sources are hosted by galaxies that have the youngest stellar population and their SMBH present the highest accretion efficient across the different AGN classes. \cite{Georgantopoulos2023} used AGN in the COSMOS field and found that highly obscured sources ($\rm N_H>10^{23}\,cm^{-2}$) live in galaxies with older stars (higher D$_n$4000 values) compared to their unobscured (or moderately obscured) counterparts. Their analysis also showed that highly obscured AGN have lower n$_{Edd}$ compared to unobscured sources. 

It is important to note, though, that our sample has significant differences compared to that used in \cite{Georgantopoulos2023}. In \cite{Georgantopoulos2023} the classification of sources is based on X-ray criteria, as opposed to the optically classified sources employed in our work. Previous works have shown that the two classification schemes do not necessarily coincide \citep[e.g.,][]{Merloni2014, Li2019, Masoura2020}. Furthermore, our dataset spans significantly lower X-ray luminosities (the majority of our sources have $\rm log\,[L_{X,2-10keV}(ergs^{-1})]<42$) compared to the luminosities probed by the COSMOS sample that is used in \citet{Georgantopoulos2023}, where $\rm 42.5<log\,[L_{X,2-10keV}(ergs^{-1})]<44.3$. Moreover, the galaxies employed in the \cite{Georgantopoulos2023} analysis are more massive compared to our galaxies, with a median difference of $\sim 0.7$\,dex. Therefore, apart from the redshift difference between the two datasets, most likely, the two studies probe different AGN populations which may have been triggered by different physical processes. Previous studies have also suggested that the comparison of the SFR of (X-ray or optically selected) obscured and unobscured AGN differs as a function of redshift and L$_X$ \citep[e.g.,][]{Mountrichas2024a, Mountrichas2024b, Mountrichas2024c}.

Our results also show higher $\lambda _{sBHAR}$ values for Sy2 galaxies compared to Composite and LINERS. Previous studies found that type 1 AGN exhibit elevated $\lambda _{sBHAR}$ values in comparison to type 2 \citep[e.g.,][]{Mountrichas2024a}. Similar outcomes have been observed when the AGN classification is based on X-ray criteria \citep[e.g.][]{Ricci2017a, Ricci2022, Georgantopoulos2023, Ricci2023, Mountrichas2024b}. The higher $\lambda _{sBHAR}$ values of type 1/unobscured AGN compared to type 2/obscured has been attributed to the effect of radiation pressure. Specifically, at higher Eddington ratios, radiation pressure may lead to a reduction in the covering factor of obscuring gas, making sources more likely to be observed as unobscured \citep{Ricci2017a}. Our analysis also incorporates CT sources, which exhibit the highest $\lambda _{sBHAR}$ values among the four AGN classes examined in this study. In Fig. 3 and the extended data Fig. 1 of \cite{Ricci2017}, there is a suggestion of elevated $\lambda _{sBHAR}$ values for CT sources compared to Compton-thin AGN ($\rm N_H=10^{22-24}\,cm^{-2}$). High $\lambda _{sBHAR}$ values for CT AGN were also reported by \cite{Brightman2016} (log\,$\lambda _{sBHAR} \sim -1$), using 12 megamaser AGN detected by $\it{NuSTAR}$.

In a study by \cite{Leslie2016}, they employed data from the SDSS data release 7, utilizing properties calculated by the MPA/JHU group. It is worth noting that their methods for computing host galaxy properties differ from our SED fitting analysis. Their investigation revealed that composite, Seyfert, and LINER galaxies are positioned below the main sequence, emphasizing the substantial impact of AGN activity in suppressing star-formation in these systems. Moreover, based on their findings LINERS have, on average, the lowest SFR and the highest M$_*$, among the different AGN populations. Our results align remarkably well with their observations.

In an investigation conducted by \cite{Kewley2006}, they focused on 85\,224 emission-line galaxies selected from SDSS and identified a significant distinction between Seyferts and LINERS, particularly in terms of their $n_{Edd}$. Their analysis indicated that LINERS tend to exhibit predominantly lower n$_{Edd}$ values compared to Sy2 galaxies. Our results align with these observations. Additionally, their investigation into the stellar populations of different AGN classes, based on the distributions of the D4000 spectral index, revealed that LINER galaxies have older stellar populations (higher D4000 values) compared to Seyferts. Once again, our findings are consistent with these outcomes.

Our results could indicate that the different types of AGN we have examined may result from distinct phases of AGN activity. For example, if a SMBH becomes active early in a galaxy's evolution, when there is plenty of gas, it might exhibit a high accretion rate, appearing as a Sy2 and potentially transitioning from CT to Sy2. On the other hand, if the AGN activity begins later in the galaxy's timeline when gas is less abundant, the SMBH would likely have a lower accretion rate, leading to a LINER \citep[e.g.,][]{TorresPapaqui2024}.

\section{Conclusions}
\label{sec_conclusions}

In this work, we used 338 galaxies at $\rm 0.02<z<0.1$ to study the AGN and host galaxy properties of different (non-QSO) AGN classes included in the SDSS-DR18 catalogue. These sources have available classification that is based on their emission-line ratios. Specifically, galaxies are  classified into Sy2, composite and LINERS.  Among these sources, we identified and select CT AGN, using their L$_X-$L$_{12\mu m}$ relation \citep{Asmus2014} and applying the threshold suggested by \cite{Pfeifle2022}. We constructed and fit the SED of the sources using the CIGALE code and applied strict criteria to include in our analysis only sources with reliable SED fitting measurements. Our sample consists of 118 Sy2, 82 composite, 124 LINERS and 14 CT sources. The 14 CT AGN are classified as Sy2 and have been excluded from the Sy2 populations. These sources have available measurements for their D$_n$4000 and H$_\delta$ spectral indices, which serve as proxies for their stellar populations. Our goal was to examine the position of these AGN populations relative to the main-sequence, compare their stellar populations and their accretion efficiency. Our main findings are summarized as follows:

\begin{itemize}

\item[$\bullet$] The majority of sources, regardless of their classification, are situated below the main-sequence. LINERS predominantly reside in galaxies characterized by higher stellar mass and lower levels of star formation activity compared to Sy2, composite, and CT sources.

\item[$\bullet$] Our findings suggest that a lower level of AGN activity corresponds to a closer alignment of the host galaxy with the main-sequence.

\item[$\bullet$] When comparing their spectral indices, LINERS exhibit the oldest stellar populations (indicated by higher D$_n$4000 values) compared to other AGN populations. Composite and Sy2 galaxies show similar stellar populations, while CT AGN host the youngest stellar populations among the classes examined in this study.

\item[$\bullet$] In LINER and composite galaxies the AGN displays the lowest accretion efficiency (lower specific black hole accretion values), while CT AGN, on average, exhibit the most efficient accretion among the four AGN populations.

\end{itemize}

In summary, our comprehensive analysis sheds light on the diverse characteristics of AGN host galaxies, emphasizing the intricate interplay between AGN activity, stellar populations, and accretion efficiency. These insights contribute to a deeper understanding of the multifaceted nature of AGN and their impact on host galaxy properties.

\begin{acknowledgements}
This project has received funding from the European Union's Horizon 2020 research and innovation program under grant agreement no. 101004168, the XMM2ATHENA project.

\end{acknowledgements}

\bibliography{mybib}
\bibliographystyle{aa}

\end{document}